\documentclass[10pt]{article}
\usepackage[OE]{express}
\usepackage{physics}
\usepackage{caption}
\usepackage[labelsep=period]{caption}

\begin{document}
\title{Experimental demonstration of a directionally-unbiased linear-optical multiport}

\author{Shuto Osawa,\authormark{1,*} David S. Simon,\authormark{1,2} and Alexander V. Sergienko\authormark{1,3,4}}

\address{\authormark{1}Department of Electrical and Computer Engineering, Boston University, 8 Saint Mary's Street, Boston, Massachusetts 02215, USA\\
\authormark{2}Department of Physics and Astronomy, Stonehill College, 320 Washington Street, Easton, Massachusetts 02357, USA\\
\authormark{3}Department of Physics, Boston University, 590 Commonwealth Avenue, Boston, Massachusetts 02215, USA\\
\authormark{4}Photonics Center, Boston University, 8 Saint Mary's Street, Boston, Massachusetts 02215, USA}
\email{\authormark{*}sosawa@bu.edu} 



\begin{abstract}
All existing optical quantum walk approaches are based on the use of beamsplitters and multiple paths to explore the multitude of unitary transformations of
quantum amplitudes in a Hilbert space. The beamsplitter is naturally a directionally biased device: the photon cannot travel in reverse direction. This causes
rapid increases in optical hardware resources required for complex quantum walk applications, since the number of options for the walking particle grows with
each step. Here we present the experimental demonstration of a directionally-unbiased linear-optical multiport, which allows reversibility of photon direction. An amplitude-controllable probability
distribution matrix for a unitary three-edge vertex is reconstructed with only linear-optical devices. Such directionally-unbiased multiports allow direct
execution of quantum walks over a multitude of complex graphs and in tensor networks. This approach would enable simulation of complex Hamiltonians of physical
systems and quantum walk applications in a more efficient and compact setup, substantially reducing the required hardware resources.
\end{abstract}

\ocis{(270.0270) Quantum Optics; (270.5585) Quantum Information and processing.} 


\section{Introduction}
Over the past decades, quantum computers have stimulated significant interest, based on their potential advantages over conventional computing devices in certain specific tasks. Several quantum algorithms have been proposed for quantum computers \cite{shor,grover}, but it is clear that significant time and resources are still required in order for general-purpose quantum computing to become a reality. It is therefore beneficial to revisit Feynman's original approach of creating special purpose quantum computing devices that are capable of efficiently simulating specific features of a complex physical system by utilizing another simple and controllable quantum device (i.e. a quantum simulator or ``analog'' quantum computer ) \cite{feynman}. Quantum walks \cite{aharonov,portugal,feldman1} on complex graphs are good candidates for executing this approach, given the statistical nature of quantum mechanical phenomena. Furthermore, quantum walks on graphs have been shown to possess the capability of performing universal computation \cite{childs}. Linear optics is one plausible physical architecture for implementing quantum walks due to its stability to decoherence, room temperature operation, and its scalability through on-chip integration.

A beam splitter (BS) \cite{Saleh} is a device of central importance in realizing any discrete quantum unitary operation \cite{Reck,Carolan}, Hadamard-coin-based quantum walks \cite{kempe}, and other quantum interferometer-based experiments \cite{Spring,Clements,Crespi,DeNicola,Wang,Peruzzo}. It is of interest to look at BS properties from the point of its symmetries. The BS has reflection and time-reversal invariance allowing the two input and two output ports to be interchanged. However, the BS is asymmetric and directionally-biased in the sense that an input photon cannot leave the same port through which it entered. This property of directional bias forces standard implementations of linear-optical quantum walks to form a unidirectional, expanding tree of possible optical paths, leading to a rapid increase in the hardware requirement needed to cover all possible probability amplitudes that are present in the Hamiltonian of the physical system under investigation \cite{Crespi,metcalf,broome}. Similarly, all the optical multiports currently in use, such as tritters, share the same directionally-biased behavior \cite{tritter,threemz}. The multiports considered here have some similarities with a previously-introduced type of nondirectional coupler known as reflective star couplers \cite{Zhang,Kogelnik}. However, unlike star couplers, the directionally unbiased multiports discussed here have relative output amplitudes at each port that can be readily tuned to desired values without dynamically changing splitting ratios of beam splitters. This additional flexibility opens up a broad range of new applications. 

\begin{figure}[htbp]
\begin{center}
\includegraphics[width=3in]{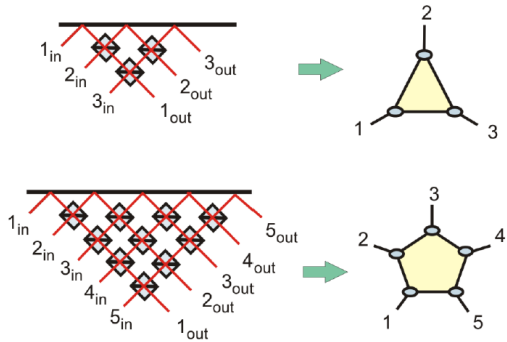}
\captionsetup{font = footnotesize}
\captionsetup{width = 0.8\textwidth}
\caption{The functioning of a linear array with $N$ input and output ports can be reproduced with a single directionally-unbiased $N$-port. The number of required beam splitters grows quadratically in the former case, but only linearly in the latter case. In many applications, the number of detectors will also increase more slowly, leading to a significant saving in resources when unbiased multiports are used.}\label{scalingfig}
\end{center}
\end{figure}

It is possible to construct a directionally-unbiased multiport by combining several linear-optical elements such as beam splitters and mirrors to form a single
compound device \cite{multiport}. Directionally unbiased $m$-ports for any $m$ can be constructed in this way. This multiport can be seen as the physical
implementation of a scattering vertex for quantum walks in undirected graph systems, with the added benefit of easy control of the unitary transformation between
input and output amplitudes. By connecting multiple copies of such devices one could carry out a variety of different experiments, such as 1D and
higher-dimensional discrete quantum walks, realizing group transformation with Bell states \cite{multiport}, simulation of physical system Hamiltonians
\cite{hamiltonian}, and experimental exploration of topological phases \cite{topo} using linear optics.
In particular, in quantum walk applications, because of direction reversibility the walk can occur along a single line, rather than perpendicular to the overall direction of motion imposed by the beam splitter. Consequently, the setup has one fewer dimension, greatly reducing the required resources and the experimental complexity. In general, the functioning of a system of beam splitters whose resources scale as $N^p$ in the number of input/output ports or time steps $N$ can be reproduced for large $N$ by a system of optical multiports scaling more slowly by one power, $N^{p-1}$. One example of this are shown in Fig. \ref{scalingfig}; a further example related to quantum walks can be seen in Ref. \cite{topo}.

In this paper we report on the first experimental demonstration of a directionally-unbiased linear-optical three-port and explore its properties.

\section{Directionally unbiased three-ports}

\begin{figure}[htbp]
\begin{center}
\includegraphics[width=3in]{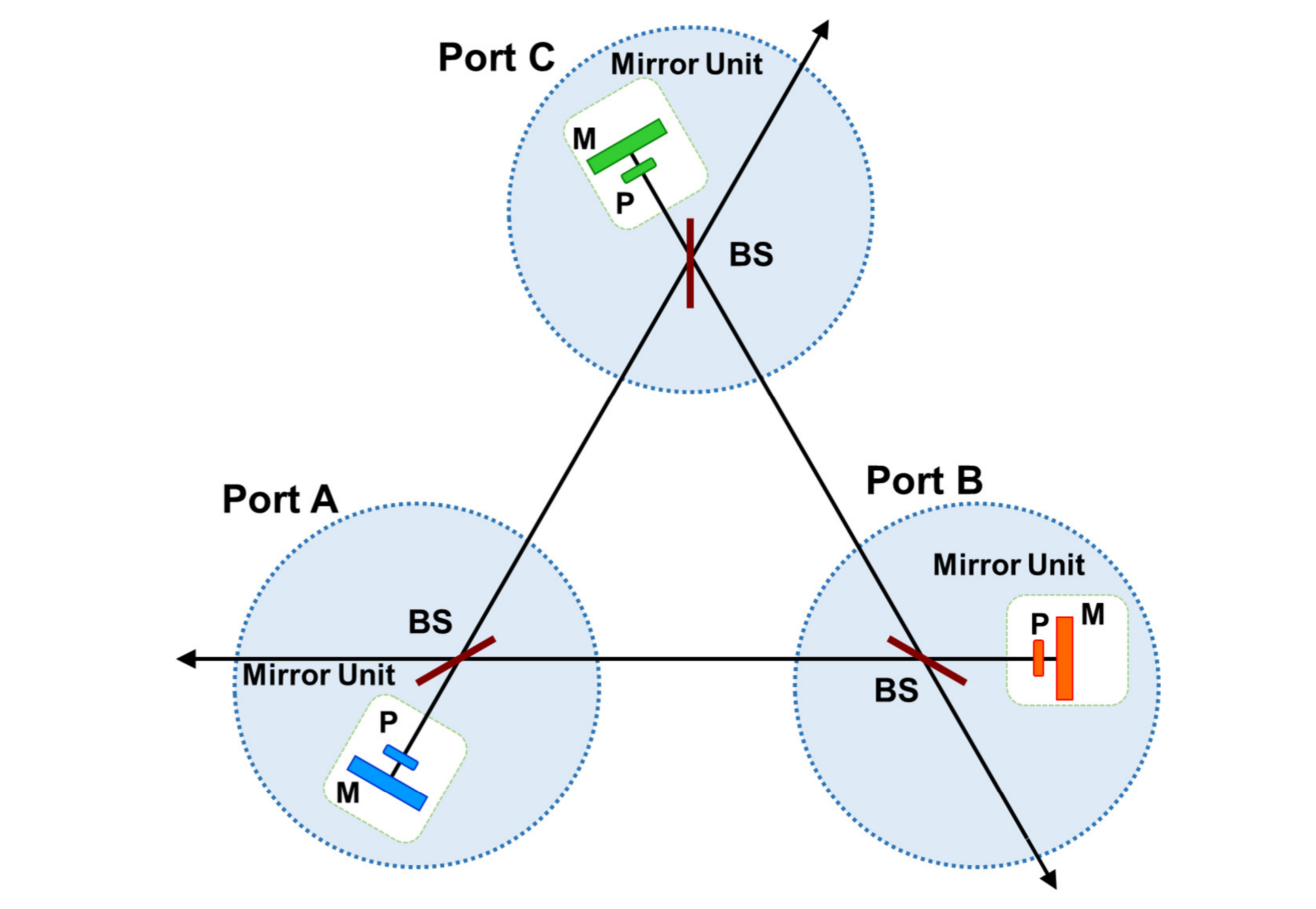}
\captionsetup{font = footnotesize}
\captionsetup{width = 0.8\textwidth}
\caption{Theoretical implementation of three-ports. Each port is labeled as port A, port B and port C. Each port is adjacent to a unit containing a non-polarizing beamsplitter (BS), a phase plate (P) and a mirror (M). A photon can enter and leave through any of the three ports.}
\end{center}
\end{figure}

We focus on the simplest case, where the number of ports is $m = 3$. A general schematic of the original theoretical idea is given in Fig. 2 \cite{multiport}. It
has three inputs/outputs labeled as port A, B, and C. The multiport internal structure consists of three beamsplitters with 50/50 transmission reflection ratio
and three mirror units with control over the phases of reflected photons. When the photon enters from port A, it is either transmitted or reflected upon hitting
the first, as well as subsequent, beam splitters. The reflected photon acquires $\frac{\pi}{2}$ phase shift at each beam splitter, with additional controllable
phase shifts from phase plates. The photon then accumulates different phases for each potential path through the system; this phase depends on the number of
encounters with beam splitters and mirror units, equivalently, on the number of edges traversed.

In the Figure, each mirror unit consists of a phase plate (P) and a mirror (M), which together impart some controllable total phase shift $\phi$ at each
encounter. The multiport input/output properties are defined by a {\em coherent summation} of amplitudes of all possible paths having the same input and output.
In order to avoid turning this into a classical, decoherent sum over probabilities, information on which internal path each photon took must remain inaccessible.
Therefore, the detection of the photon output probability at ports A, B and C must be performed over a time much longer than the transit time between beam
splitters in order to avoid paths to be distinguished by exit time. After a sufficiently long time, the output probability amplitudes for the photon to exit at
the three ports after entering the multiport at port A are given by:

\begin{equation}
\begin{aligned}
A \rightarrow A = {} & \frac{1}{4}e^{i\phi_C}+\frac{1}{4}e^{i\phi_B}-i\frac{1}{8}e^{i(\phi_B+\phi_C)}-i\frac{1}{8}e^{i(\phi_B+\phi_C)}+\frac{1}{16}e^{i(\phi_A+\phi_B+\phi_C)}+\frac{1}{16}e^{i(\phi_A+\phi_B+\phi_C)}\\
&-\frac{1}{16}e^{i(\phi_B+\phi_C+\phi_B)}-\frac{1}{16}e^{i(\phi_C+\phi_B+\phi_C)}-\frac{1}{16}e^{i(\phi_B+\phi_A+\phi_B)}-\frac{1}{16}e^{i(\phi_C+\phi_A+\phi_C)} + \dots
\end{aligned}\label{transitionA}
\end{equation}

\begin{equation}
\begin{aligned}
A \rightarrow B = {} & i\frac{1}{2}-\frac{1}{4}e^{i\phi_C}-i\frac{1}{8}e^{i(\phi_A+\phi_B)}+i\frac{1}{8}e^{i(\phi_A+\phi_C)}+i\frac{1}{8}e^{i(\phi_B+\phi_C)}+\frac{1}{16}e^{i(\phi_A+\phi_B+\phi_C)}\\
&-\frac{1}{16}e^{i(\phi_A+\phi_B+\phi_C)}-\frac{1}{16}e^{i(\phi_A+\phi_B+\phi_C)}+\frac{1}{16}e^{i(\phi_C+\phi_B+\phi_C)}+\frac{1}{16}e^{i(\phi_C+\phi_A+\phi_C)} +\dots
\end{aligned}
\end{equation}\label{transitionB}

\begin{equation}
\begin{aligned}
A \rightarrow C = {} & i\frac{1}{2}-\frac{1}{4}e^{i\phi_B}+i\frac{1}{8}e^{i(\phi_A+\phi_B)}-i\frac{1}{8}e^{i(\phi_A+\phi_C)}+i\frac{1}{8}e^{i(\phi_B+\phi_C)}+\frac{1}{16}e^{i(\phi_A+\phi_B+\phi_C)}\\
&-\frac{1}{16}e^{i(\phi_A+\phi_B+\phi_C)}-\frac{1}{16}e^{i(\phi_A+\phi_B+\phi_C)}+\frac{1}{16}e^{i(\phi_B+\phi_C +\phi_B)}+\frac{1}{16}e^{i(\phi_B+\phi_A+\phi_B)} +\dots ,
\end{aligned}
\end{equation}\label{transitionC}

where $\phi_A,\phi_B,\phi_C$ are the total phases acquired by the photon from mirror units A, B, C, respectively, and a $\frac{\pi}{2}$ phase is acquired from each beamsplitter reflection. Small additional terms of the order $\left(\frac{1}{2}\right)^5$ or higher have been left out from consideration. Similar equations can be obtained for inputs B and C. Here, we have also assumed that phase shifts gained from propagation between mirror units are integer multiples of $2\pi$, so that they play no role.

After a sufficient number of beamsplitter encounters, the output amplitudes converge to fixed values \cite{multiport}. Let $N$ to be the number of beamsplitter encounters for the photon before exiting the multiport or, equivalently, the number of time steps, where the unit of time is the travel time between consecutive beamsplitter encounters. The cumulative probability to exit a multiport as a function of $N$ is given in table 1. One could observe that after N = 8 in over 99 percent of trials of the photons have left the system, as indicated in the Table 1.
\begin{table}
\captionsetup{justification = centering}
\captionsetup{font = {footnotesize,bf}}
\captionsetup{width = 0.8\textwidth}
\caption{Probability amplitudes for the port A, B and C, exit probability in respect to a specific numbers of BS encounters, and the cumulative exit probability are shown. Columns A exit, B exit, and C exit represent exit probability amplitude at each port with initial photon entering at port A. Exit Probability shows the exit probability for each possible number of beamspliter hits. The cumulative exit probability is obtained from summing all the exit probabilities. As N increases, the cumulative exit probability quickly approaches 1 and a photon leaves the unit.} \label{tab:sometab}
\centering

  \begin{tabular}{|c|c|c|c|c|c|c|}
  \hline
    N & A exit & B exit & C exit & Exit Probability & Cumulative Exit Probability\\
    2 & 0 & $\frac{i}{2}$ & $\frac{i}{2}$ & $\frac{1}{2}$ & 0.5\\
    4 & $\frac{-i}{2}$ & $\frac{i}{4}$ & $\frac{i}{4}$  & $\frac{3}{8}$ & 0.875\\
    6 & $\frac{i}{4}$ & $\frac{-i}{8}$ & $\frac{-i}{8}$ & $\frac{3}{32}$ & 0.96875\\
    8 & $\frac{-i}{8}$ & $\frac{i}{16}$ & $\frac{i}{16}$ & $\frac{3}{128}$ & 0.99219\\
    10 & $\frac{i}{16}$ & $\frac{-i}{32}$ & $\frac{-i}{32}$ & $\frac{3}{512}$ & 0.99805\\
    \hline
  \end{tabular}

\end{table}

A transfer matrix expression for the multiport can be obtained by substituting specific phase settings in Eq. (1), (2), and (3). We set $\phi_A$ = $\phi_B$ = $\phi_C$ = $\frac{\pi}{6}$ in this paper to explore the most symmetric case that would make this unit suitable for quantum walks on unstructured graphs. Carrying out an exact summation of the series in Eqs. (1) - (3), the resulting expected transfer matrix is:

\begin{equation}
U
=
\begin{pmatrix}
U_{A \rightarrow A} & U_{A \rightarrow B}& U_{A \rightarrow C}\\
U_{B \rightarrow A}& U_{B \rightarrow B}& U_{B \rightarrow C}\\
U_{C \rightarrow A}& U_{C \rightarrow B}& U_{C \rightarrow C}
\end{pmatrix}
 = \frac{1}{\sqrt{3}}e^{\frac{2\pi i}{3}}
\begin{pmatrix}
e^{\frac{-2\pi i}{3}} & 1&1\\
1&e^{\frac{-2\pi i}{3}}&1\\
1&1&e^{\frac{-2\pi i}{3}}
\end{pmatrix}
\end{equation}

The transfer matrix acts on an input state via the transformation:
$\ket{\psi_{output}}=U\ket{\psi_{input}}$.
Squaring the modulus of each matrix element gives the final transition probability distribution that should be observed in experiment.

\begin{equation}
P =
\begin{pmatrix}
|U_{A \rightarrow A}|^2 & |U_{A \rightarrow B}|^2&|U_{A \rightarrow C}|^2\\
|U_{B \rightarrow A}|^2&|U_{B \rightarrow B}|^2&|U_{B \rightarrow C}|^2\\
|U_{C \rightarrow A}|^2&|U_{C \rightarrow B}|^2&|U_{C \rightarrow C}|^2
\end{pmatrix}
=\frac{1}{3}
\begin{pmatrix}
1&1&1\\
1&1&1\\
1&1&1
\end{pmatrix}
\end{equation}

\section{Experiment}

The operation of a directionally-unbiased symmetric three-edge vertex device has been demonstrated by characterizing probability distribution matrix elements responsible for all possible photon input/output transitions. The actual experimental setup is shown in Fig. 3. A continuous-wave 10 mW laser operating at 633 nm (NECSEL single longitudinal mode) with a very long coherence length (> 1 km) was used. This condition ensures the coherent superposition requirement for all possible photon paths inside the unit. The beamsplitters have 50/50 transmission/reflection ratio at 633 nm. The phase shifts imparted by the mirror units are controlled by piezo actuators on translation stages. The input beam was attenuated to a single-photon level with polarizers. The multiport arrangement indicated in Fig. 3 is for the input at port A and outputs from A, B and C.

\begin{figure}[htb]
\begin{center}
\includegraphics[width=4in]{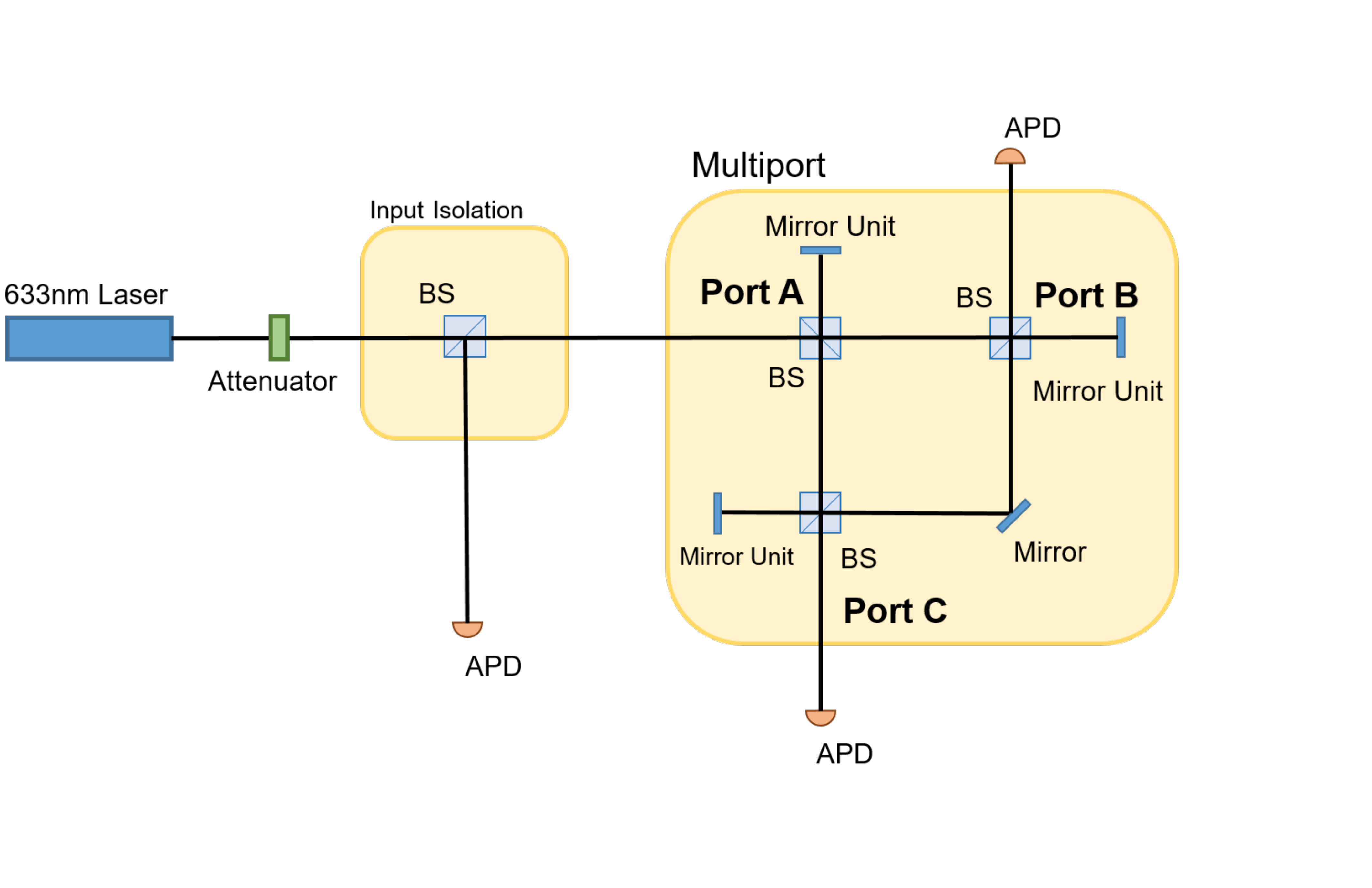}
\vspace{-0.3in}
\captionsetup{font = footnotesize}
\captionsetup{width = 0.8\textwidth}
\caption{Experimental setup for detecting probabilistic single-photon distribution in a three-port with a photon entering at the port A. A laser with long coherence length (> 1 km) operates at 633 nm. The input beam is attenuated to a single-photon level prior entering the multiport. The input isolation allows decoupling the input photon form the one exiting through the same port A. This multiport consists of three beamsplitters, three mirrors with piezo actuators to control internal phases, and an extra mirror in one arm. The number of exiting photons is counted by a single-photon avalanche diode modules (APD). Phases in each branch of the multiport are set to $\frac{\pi}{6}$ using mirror units.}
\end{center}
\end{figure}

This corresponds to the first matrix row in Eq. (5). The beam splitter at the port A serves as a 50\% isolation device separating half of the output photons statistically into a different path for detection. An optical circulator would be a better device to use in the future in order to separate all of the outgoing photons from the incoming ones. A similar setup is used to measure transitions with input at port B and port C. This would correspond to rows 2 and 3 in Eq. (5), respectively. The  outgoing photons at each port were counted using fiber-coupled single-photon avalanche diodes (PerkinElmer SPCM-AQR-15-FC), as indicated in the Fig. 3. The real experimental configuration is equivalent to the theoretical schematic depicted in Fig. 2, despite the slightly different shape; this more complex experimental arrangement is necessary in the current case because the equidistant triangular setup described in section 2 cannot be easily realized with cube beam splitters.  The distance between beam splitters A-B and A-C is set to be identical, while B-C is set to be twice as long as A-B and has one additional mirror. This does not affect the overall probability distribution in the matrix Eq. (4) when the photon coherence is sufficiently long. What is important is the  phase balance in all possible paths inside the multiport. The mirror units are translated for an additional  phase $\pi$ to recover the balance due to one extra mirror presence. Alternatively, this extra $\pi$ phase can be compensated by arranging the structure so that all three path have the same numbers of mirrors. The same result could be achieved by inserting a phase plate in the path with an extra mirror.

 \begin{figure}[htb]
\centering
\includegraphics[width=4in]{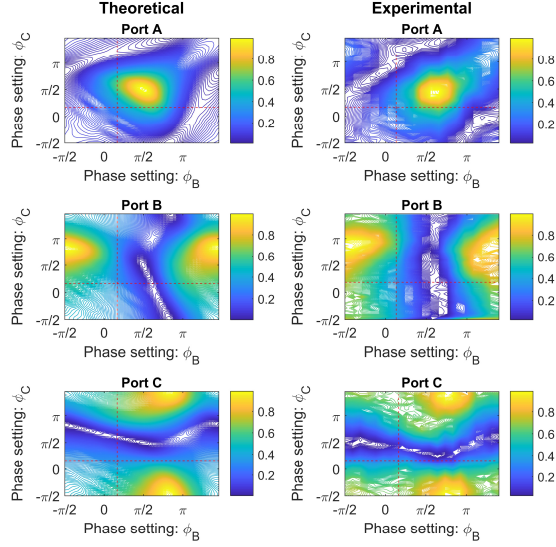}
\captionsetup{font = footnotesize}
\captionsetup{width = 0.8\textwidth}
\caption{Theoretical simulation (left) and experimental results (right) at $\phi_A$ = $\frac{\pi}{6}$.The contour graphs represent probability distribution for the input-output combinations  A-A, A-B, and A-C where the minimum is close to 0 and maximum is close to 1. Light color coding (yellow) corresponds to a high probability and dark color (blue) indicates low probability. The phase at the mirror unit A is set to $\phi_A=\frac{\pi}{6}$ for all three plots . The left contour graphs are the theoretical simulation. Dashed lines are drawn at $\phi_B = \frac{\pi}{6}$ and $\phi_C = \frac{\pi}{6}$. The cross section indicates the point when all three phases ($\phi_A$,$\phi_B$,$\phi_C$) = ($\frac{\pi}{6}$,$\frac{\pi}{6}$,$\frac{\pi}{6}$). The contour graphs in the right column present results of experimental observation. The dashed lines are drawn to indicate phase settings at $\frac{\pi}{6}$. The phase step in the contour graph is $\frac{\pi}{15}$.}
\end{figure}

The weak photon source statistics follow a Poisson distribution, and the input coherent laser beam is attenuated to the single-photon level, meaning its average photon number during the detection interval must be smaller than 1. Our single-photon avalanche photodetector integration time is 35 ns. The average photon number within the detector integration window is below 0.1 when the input beam is attenuated to 1 nW average power. The detector quantum efficiency is 0.6 at the 633 nm wavelength of the source.

The system under consideration could be considered as a coherent superposition of several interferometers with three independent phase controls. The detected outcome at each port forms an interference pattern of intensity as a function of those phases. The quality of the multiport alignment is verified by determining the visibility of such interference patterns:

\begin{equation}
V = \frac{n_{max}-n_{min}}{n_{max}+n_{min}} ,
\end{equation}

Where $n_{max}$ is the maximum number of detected photons, $n_{min}$ is the minimum number of detected photons.

The problem of setting all three phases to the same value is not very straightforward, since the phases are relative parameters. Interference contour maps were generated to cover all possible phases from 0 to $2\pi$ in order to find a position when all three of them have the same $\frac{\pi}{6}$ value (see Fig. 3).

The distance between consecutive beamsplitter hits in the multiport is a bit less than 30 cm. Any incident photon will exit the system with a probability 0.998 after N = 10 interactions \cite{multiport}, so one has to insure a coherent (indistinguishable) travel of the photon over about 300 cm in this case. This is significantly shorter than the coherence length of the source used in our experiment (> 1 km). One should point out that dividing this number by the speed of light we obtain an estimate for a maximum time any photon would stay inside the multiport (about 10 ns).

The optical table was passively stabilized with active mechanical vibration isolation. By this means, the required interferometric stability of operation was achieved over a 5 hour period, which was sufficient for automatic scans of all required high-visibility interference patterns illustrated in Fig. 4. In future free-space experiments requiring greater data acquisition times, active stabilization loops may be needed. Eventual transfer of this configuration into a waveguided on-chip design (using currently available technology of making polarization insensitive laser-written waveguides in silica glass) will help significantly to enhance performance stability and transverse alignment, and to reduce the long-term longitudinal phase drift. This offers extensive opportunities for miniaturizing the device on a chip scale, allowing greater compactness and stability, as well as reducing the coherence requirements.

Multiple contour maps were obtained by translating two mirror units B and C (one at a time) when the phase from a mirror unit $\phi_A$ is fixed at a certain value. Then the procedure was repeated multiple times for new values of $\phi_A$. A subset of such contour maps corresponding to the situation when the photon is inserted at port A is illustrated in Fig. 4. The situation when the photon enters at ports B or C generates a similar set of contour plots.  The comparison of such experimental contour maps with those obtained in theoretical simulation enables one to identify a point corresponding to $\frac{\pi}{6}$ phase shift value in all three phases (indicated by a dashed line). The final probability distribution for single photons to enter and exit any of the ports has been performed at these particular phase settings from all three mirror units A, B, and C.

In order to compare to the real experimental situation, Eqs. (1) - (3) must be modified to account for the additional mirror present (see Fig. 3). This is done by shifting each phase acquired from the mirror units by $\pi$. The resulting equations will have the following modifications: $\phi_A \rightarrow \phi_A + \pi$, $\phi_B \rightarrow \phi_B + \pi$, and $\phi_C \rightarrow \phi_C + \pi$.
%
%
%
Under these conditions $A \rightarrow A$ changes sign, $A \rightarrow B$ and $A \rightarrow C$ remain the same, with similar changes for other input ports.  The theoretical unitary transfer matrix corresponding the current experimental configuration is now:

\begin{equation}
 U = \frac{1}{\sqrt{3}}e^{\frac{2\pi i}{3}}
\begin{pmatrix}
-e^{\frac{-2\pi i}{3}} & 1&1\\
1&-e^{-\frac{2\pi i}{3}}&-1\\
1&-1&-e^{-\frac{2\pi i}{3}}
\end{pmatrix}
\end{equation}

This insignificant phase modification in the unitary transfer matrix Eq. (7) is a reflection of the need to use several bulk optical elements (such as beam splitters and mirrors) during the first experimental implementation. (Currently available technology would allow the original theoretically-formulated unitary transfer matrix indicated in Eq. (4) to be realized using a waveguided platform an integrated configuration that executes all features of the multiport in Fig. 2. Such an on-chip configuration would be highly desirable due to improved compactness and stability.)  This {\em phase modification} leaves the final {\em probability distribution} for the transfer matrix {\em unchanged} from that of section 2:

\begin{equation}
P=\frac{1}{3}
\begin{pmatrix}
1&1&1\\
1&1&1\\
1&1&1
\end{pmatrix}
\end{equation}

\section{Experimental results}

The quality of the overall multiport alignment has been verified by observing a single photon interference at each port by sequentially feeding a single-photon input state into each of ports A, B, and C.  The experimentally observed visibility is calculated for all 9 possible outcomes corresponding to a photon state inserted in port A,  port B, and port C. Three plots corresponding to insertion at Port A case are illustrated in Fig. 4 (right column). An average observed visibility of $0.97\pm0.01$ was achieved. A phase step of $\pi\over{15}$ was taken between each point on the plots. All presented data are based on direct observation, with no background subtraction performed. A small reduction in the visibility came from the detector dark counts, slight reflection on the beamsplitter surfaces, and from minor imperfection of beamsplitter surface alignment that results in the imperfect overlap of the beams. All experimental probability detection maps for each port in Fig. 4 are normalized by their maxima to obtain a probability distribution. The contour map in the right column of Fig. 4 represents the photon count distribution for the fixed $\phi_A$ = $\frac{\pi}{6}$ and phase sweeps from 0 to $2\pi$ for $\phi_B$ and $\phi_C$ by changing the voltages on piezo actuators responsible for the position of corresponding metal mirrors in space. Each map is compared with its theoretical simulations in the left column in Fig. 4. The same method and similar contour maps were used for detecting outcome probabilities when a single photon has been inserted in B and C. The probability distribution for the photon to transition between any pair of ports has been reconstructed by statistically averaging the raw data over 1.8 s.

The final reconstructed probability matrix at ($\phi_A$,$\phi_B$,$\phi_C$) = ($\frac{\pi}{6}$,$\frac{\pi}{6}$,$\frac{\pi}{6}$) is given by:

\begin{equation}
P_{exp} =
\begin{pmatrix}
0.332\pm0.007 & 0.343\pm0.007&0.340\pm0.004\\
0.340\pm0.005&0.339\pm0.006&0.328\pm0.008\\
0.332\pm0.005&0.340\pm0.012&0.338\pm0.007
\end{pmatrix}
\end{equation}


Losses generally play a negative role in quantum optics experiments. However, the role of losses in the case considered here is not significant. The device is probabilistic and operates with one photon at a time in the system. The loss of a photon simply results in discarding this particular trial from detection and recording of all possible outcomes. This does not affect the quality of the next trial and does not degrade the overall visibility of the detected outcome. The effect of losses shows up simply as an increase in the required accumulation time needed to get sufficient statistics when characterizing the multiport. The probability distribution (input-output) matrix elements were characterized one by one and are independent of one another.  These are obtained from a contour map using visibility as the major quality parameter. The loss does not affect the visibility normalization process since we have the same loss across the contour map.

The interference patterns observed at each port depend on the fact that all of the possible paths within the multiport remain coherent with each other. The high visibility observed in each output port is therefore indicates a high level of mutual coherence of the output at the different ports, a requirement for multi-photon (and especially entangled-photon) quantum information processing applications.

\section{Discussion}

The experimentally  observed probability distribution for the directionally-unbiased linear-optical multiport illustrates the validity of the original theoretical concept \cite{multiport} offering to construct coherent multi-edged vertices that are suitable for experimentally executing universal quantum walks on arbitrary graphs. The natural reversibility of the photon flow in such multiports offers a dramatic reduction in the amount of required hardware resources when compared with currently exploited systems based on beamsplitter trees. The encounter of such multiport during a quantum walk procedure could  be considered as a quantum coin application. The greater number of edges $(N \geqslant2)$ at every application of such quantum coin speeds up significantly the coverage of higher dimensions in Hilbert spaces. This opens new possibilities when simulating dynamic Hamiltonians of complex physical systems.
\begin{figure}[htbp]
\centering
\includegraphics[width=5in]{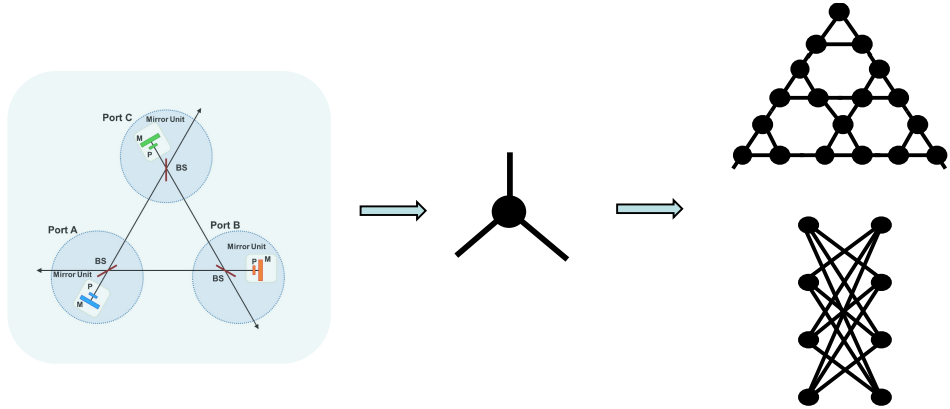}
\vspace{-0.75in}
\captionsetup{font = footnotesize}
\captionsetup{width = 0.8\textwidth}
\caption{Directionally-unbiased linear optical three-edge vertex device demonstrated in this paper could be used for building complex
quantum graphs. The efficiency of approaches based on quantum walks on graphs hold promise of addressing some complex scientific and technical problems
more efficiently than conventional numerical computational methods. For example, networks such as the one on the bottom right, which can be implemented with
unbiased three-ports, are
used in machine learning applications.}
\label{graphsfig}\end{figure}

This multiport could find a number of applications in the areas of quantum information processing and quantum simulation of dynamic Hamiltonians.  The quantum walk on a specially designed network of such three-edge multiports offers configuration flexibility enabling  to simulate Hamiltonians of complex polymer chains, energy band structure of semiconductor materials including topological insulators  \cite{hamiltonian}. Although an unbiased three-port was examined here, $m$-ports with $m>3$ can be constructed in a similar manner.

The main result is shown for a totally symmetric three-edge vertex multiport. This multiport design  allows to realize a multitude of different unitary
transformation matrices (see \cite{multiport}) without ever changing beamsplitter splitting ratios. A simple modification in phase shifts at each of the mirror
units can modify the transfer matrix amplitudes, allowing output distributions to be controllably tailored for different purposes and opening additional
possibilities for simulating physical systems with a range of output probability distributions. This feature could be of particular interest for building quantum
optical machine learning units that require dynamic adjustment of transition weights during the learning process (Fig. \ref{graphsfig}).

Connecting multiple copies of a three-edge multiports could result in practical realization of a quantum walk on several complex graphs with different
topological properties, using only linear optics. A 1D quantum particle walk on a lattice of such multiport structures corresponding to an SSH-type system has
been proposed, offering conditions to demonstrate topological structures and distinguishable winding numbers \cite{topo}. A 2D distribution of three-edge
vertices (see Fig. 5) has even more interesting applications in the area of designing novel types of topological insulators, quantum walk over fractal states
such as Sierpinski gasket \cite{Crownover}, and serve as natural elements in constructing tensor networks \cite{Werner}.
The quantum character of signal processing in tensor networks promises significant speed up in the problem solving large systems of differential equations
\cite{Bachmayr}. It would be practically impossible to execute quantum walks on complex graph structures of reasonable size using conventional
binary coins realized with optical beam splitters; directionally unbiased $m$-ports provide a much more feasible route to implementing such structures while offering significant savings in resources.

Multi-photon states can also be inserted instead of single photon inputs. The use of multi-photon input states more easily allows one to fully reconstruct a
transfer matrix by explicitly recovering the phase information as well. In addition, such a three-port device with two-photon Bell-state inputs could be used to
navigate a group structure when a pair of correlated Bell-states are injected into a non-overlapping set of ports \cite{multiport}.

In conclusion, we have demonstrated experimentally the operation of a directionally-unbiased linear-optical three-port unitary device at a particular phase setting of $\frac{\pi}{6}$ by reconstructing a probability distribution of all possible propagation patterns with an interferometric method. The effective operation of the device as a symmetric three-edge vertex suitable for quantum walks on graphs has been demonstrated. This provides one more step in the  goal of using such directionally-unbiased multiports to implement complex quantum walks on graphs and in ultimately demonstrating their practical use in achieving quantum speedup for the solution of specific physical simulations.

\section*{Funding}

National Science Foundation EFRI-ACQUIRE (ECCS-164096); AFOSR Grant (FA9550-18-1-0056); Northrop
Grumman NG Next.


\begin{thebibliography}{99}
\bibitem{shor} P. W. Shor, ``Algorithms for quantum computation: discrete logarithms and factoring,'' in \textit{Proceedings
of the Symposium on the Foundations of Computer Science, Los Alamitos, California} (IEEE Computer Society,
1994), pp. 124--134
\bibitem{grover} L. K. Grover, ``A fast quantum mechanical algorithm for database search,'' in \textit{28th Annual ACM Symposium on Theory of Computing} (ACM Press, 1996), pp. 212--219.
\bibitem{feynman} R. P. Feynman, ``Simulating physics with computer,'' Int. J. Theor. Phys. \textbf{21}, 467--488 (1982).
\bibitem{aharonov} Y. Aharonov, L. Davidovich, and N. Zagury, ``Quantum random walks,'' Phys. Rev. A \textbf{48}, 1687--1690 (1993).
\bibitem{portugal} R. Portugal, \textit{Quantum walks and search algorithms} (Springer Science \& Business Media, 2013).
\bibitem{feldman1} E. Feldman and M. Hillery, ``Scattering theory and discrete-time quantum walks,'' Phys. Lett. A \textbf{324}(4), 277--281 (2004).
\bibitem{childs} A. M. Childs, ``Universal computation by quantum walk,'' Phys. Rev. Lett. \textbf{102}(18), 180501 (2009).
\bibitem{Saleh} M. C. Teich and B. E. Saleh, \textit{Fundamentals of photonics} (Wiley, 2013). 
\bibitem{Reck} M. Reck, A. Zeilinger, H. J. Bernstein, and P. Bertani, ``Experimental realization of any discrete unitary operator,'' Phys. Rev. Lett. \textbf{73}(1), 58 (1994).
\bibitem{Carolan} J. Carolan, C. Harrold, C. Sparrow, E. Mart\'in-L\'opez, N. J. Russell, J. W. Silverstone, P. J. Shadbolt, N. Matsuda, M. Oguma, M. Itoh, G. D. Marshall, M. G. Thompson, J. C. F. Matthews, T. Hashimoto, J. L. O'Brien, and A. Laing, ``Universal linear optics,'' Science \textbf{349}(6249), 711--716 (2015).
\bibitem{kempe} J. Kempe, ``Quantum random walks-an introductory overview,'' Contemp. Phys. \textbf{44}(4), 307--327 (2003).
\bibitem{Spring} J. B. Spring, B. J. Metcalf, P. C. Humphreys, W. S. Kolthammer, X. Jin, M. Barbieri, A. Datta, N. Thomas-Peter, N. K. Langford, D. Kundys, J. C. Gates, B. J. Smith, P. G. R. Smith, and I. A. Walmsley, ``Boson sampling on a photonic chip,'' Science \textbf{339}, 798--801 (2012).
\bibitem{Clements} W. R. Clements, P. C. Humphreys, B. J. Metcalf, W. S. Kolthammer, and I. A. Walmsley, ``Optimal design for universal multiport interferometers,'' Optica \textbf{3}(12), 1460--1465 (2016).
\bibitem{Crespi} A. Crespi, R. Osellame, R. Ramponi, D. J. Brod, E. F. Galvao, N. Spagnolo, C. Vitelli, E. Maiorino, P. Mataloni, and F. Sciarrino, ``Integrated multimode interferometers with arbitrary designs for photonic boson sampling,'' Nature Photon. \textbf{7}, 545--549 (2013).
\bibitem{DeNicola} F. D. Nicola, L. Sansoni, A. Crespi, R. Ramponi, R. Osellame, V. Giovannetti, R. Fazio, P. Mataloni, and F. Sciarrino, ``Quantum simulation of bosonic-fermionic noninteracting particles in disordered systems via a quantum walk,'' Phys. Rev. A \textbf{89}(3), 032322 (2014).
\bibitem{Wang} J. Wang, S. Paesani, Y. Ding, R. Santagati, P. Skrzypczyk, A. Salavrakos, J. Tura, R. Augusiak, L. Man\~cinska, D. Bacco, D. Bonneau, J. W. Silverstone, Q. Gong, A. Ac\'in, K. Rottwitt, L. K. Oxenl\o we, J. L. O'Brien, A. Laing, and M. G. Thompson, ``Multidimensional quantum entanglement with large-scale integrated optics,'' Science \textbf{360}, 285 (2018).
\bibitem{Peruzzo} A. Peruzzo, M. Lobino, J. C. F. Matthews, N. Matsuda, A. Politi, K. Poulios, X. Zhou, Y. Lahini, N. Ismail, K. W\"orhoff, Y. Bromberg, Y. Silberberg, M. G. Thompson, and J. L. O'Brien, ``Quantum walks of correlated photons,'' Science \textbf{329}(5998), 1500--1503 (2010).
\bibitem{metcalf} B. J. Metcalf, N. Thomas-Peter, J. B. Spring, D. Kundys, M. A. Broome, P. C. Humphreys, X. Jin, M. Barbieri, W. S. Kolthammer, J. C. Gates, B. J. Smith, N. K. Langford, P. G. R. Smith and I. A. Walmsley, ``Multiphoton quantum interference in a multiport integrated photonic device,'' Nat. Commun. \textbf{4}, 1356 (2013).
\bibitem{broome} M. A. Broome, A. Fedrizzi, B. P. Lanyon, I. Kassal, A. Aspuru-Guzik, and A. G. White, ``Discrete single-photon quantum walks with tunable decoherence,'' Phys. Rev. Lett. \textbf{104}(15), 153602 (2010).
\bibitem{tritter} N. Spagnolo, C. Vitelli, L. Aparo, P. Mataloni, F. Sciarrino, A. Crespi, R. Ramponi, and R. Osellame, ``Three-photon bosonic coalescence in an integrated tritter,'' Nat. commun. \textbf{4}, 1606 (2013).
\bibitem{threemz} G. Weihs, M. Reck, H. Weinfurter, and A. Zeilinger, ``All-fiber three-path Mach-Zehnder interferometer,'' Opt.
Lett. \textbf{21}(4), 302--304 (1996).
\bibitem{Zhang} J. Zhang, A. B. Sharma, Y. Ni, and Z. Li, ``Investigation into network architecture and modulation scheme for MIL-STD-1773 optical fiber data buses,'' Aircraft Eng. Aero. Technol. \textbf{72}(2), 126--137 (2000).
\bibitem{Kogelnik} A. A. M. Saleh and H. Kogelnik, ``Reflective single-mode fiber-optic passive star couplers,'' J. Lightwave Technol. \textbf{6}(3), 392--398 (1988).
\bibitem{multiport} D. S. Simon, C. A. Fitzpatrick, and A. V. Sergienko, ``Group transformations and entangled-state quantum gates with directionally-unbiased linear-optical multiports,'' Phys. Rev. A \textbf{93}(4), 043845 (2016).
\bibitem{hamiltonian} D. S. Simon, C. A. Fitzpatrick, S. Osawa, and A. V. Sergienko, ``Quantum simulation of discrete-time hamiltonians using directionally-unbiased linear optical multiports,'' Phys. Rev. A \textbf{95}(4), 042109 (2017).
\bibitem{topo} D. S. Simon, C. A. Fitzpatrick, S. Osawa, and A. V. Sergienko, ``Quantum simulation of topologically protected states using directionally unbiased linear-optical multiports,'' Phys. Rev. A \textbf{96}(1), 013858 (2017).
\bibitem{Crownover} R. M. Crownover, \textit{Introduction to fractals and chaos} (Jones \& Bartlett Pub, 1995).
\bibitem{Werner} A. H. Werner, D. Jaschke, P. Silvi, M. Kliesch, T. Calarco, J. Eisert, and S. Montangero, ``Positive tensor network approach for simulating open quantum many-body systems,'' Phys. Rev. Lett. \textbf{116}(23), 237201 (2016).
\bibitem{Bachmayr} M. Bachmayr, R. Schneider, and A. Uschmajew, ``Tensor networks and hierarchical tensors for the solution of high-dimensional partial differential equations,'' Found. Comput. Math. \textbf{16}(6), 1423--1472 (2016).
\end{thebibliography}
\end{document}